\documentclass[a4paper,11pt]{article}
\usepackage{jinstpub}
\usepackage{graphicx}
\usepackage{placeins}
\usepackage{svg}
\usepackage{url}

\usepackage{subcaption}
\usepackage{siunitx}
\title{\boldmath Towards 6D Tracking: A Study Of Using Fast-Timing For Measuring Track Position, Time, And Angles}

\author[a,b,1]{Victor Turbiner\note{Corresponding author.}}
\author[c]{Elena Villhauer}
\author[a]{Saveer Jain}
\author[a]{Abhiraj Gupta}
\author[d]{Michael Cardiff}
\author[b]{Julie Segal}
\author[b]{Christopher J. Kenney}
\author[a]{Mark Horowitz}
\author[b]{Ariel Schwartzman}
\author[b]{Angelo Dragone}

\affiliation[a]{Stanford University Department Of Electrical Engineering}
\affiliation[b]{SLAC National Accelerator Laboratory}
\affiliation[c]{Enrico Fermi Institute, University of Chicago}
\affiliation[d]{Brandeis University}

\emailAdd{victur03@slac.stanford.edu}

\abstract{Next-generation particle trackers will incorporate precision timing with resolutions of tens of picoseconds. Using two-dimensional TCAD simulations of Low-Gain Avalanche Diodes (LGADs) with a \qty{25}{\micro\meter} pitch and \qty{75}{\micro\meter} thickness, we show that tilted tracks traversing multiple pixels produce systematic inter-pixel timing differences of hundreds of picoseconds. These differences must be corrected for in cluster hit time reconstruction, but they also encode the track angle. We derive an analytic linear model relating inter-pixel timing differences to the track angle, and determine the sign of the angle, which the cluster shape cannot provide, using an amplitude-weighted method. For tracks crossing at least three pixels, this enables single-layer angle reconstruction with a resolution of $2^\circ-3^\circ$ and, after a simple numerical inversion, a bias below $1^\circ$. The dominant limitation arises from large Landau fluctuations: non-uniform charge deposition along the track induces inter-pixel crosstalk that distorts the timing signals. Affected tracks can be identified and rejected using the total cluster charge. Neural networks that additionally exploit nonlinear correlations and pixel charge improve the resolution to approximately $1.7^\circ$, indicating that the computationally inexpensive linear model captures most of the available angular information. These results provide a path toward 6D tracking, in which each detector layer measures track position, time, and angles.}


\arxivnumber{2605.24736}

\begin{document}
\maketitle
\flushbottom

\section{Introduction}

Precision timing capabilities are becoming an essential feature of particle tracking detectors. The ATLAS High-Granularity Timing Detector (HGTD)~\cite{CERN-LHCC-2020-007} and CMS MIP Timing Detector (MTD)~\cite{Butler:2019rpu} for the High-Luminosity Large Hadron Collider (HL-LHC) achieve timing resolutions below \qty{50}{\pico\second}, while tracking systems proposed for future colliders---including muon colliders~\cite{bartosik2022simulateddetectorperformancemuon} and multi-TeV hadron machines~\cite{FCC_hh_2019}---will require comparable or superior temporal resolution with pixel dimensions on the order of tens of microns.

Achieving this high timing resolution while maintaining stringent power budgets requires specialized sensor technology. Low-Gain Avalanche Diodes (LGADs) \cite{FIRST-LGAD-PAPER} address these requirements by incorporating an internal charge multiplication layer that provides moderate gain\footnote{Typically $10\times$ to $30\times$}, thereby relaxing the noise and power requirements of the front-end electronics.

In this work, we investigate the temporal response of LGAD detectors to non-normal (tilted) particle tracks. When a charged particle traverses the detector at sufficiently oblique angles, it deposits energy across multiple adjacent pixels. Through Technology Computer-Aided Design (TCAD) simulations, we demonstrate that such multi-pixel events produce systematically different timing measurements across the traversed pixels, with variations reaching hundreds of picoseconds. Given that state-of-the-art timing detectors target resolutions of tens of picoseconds, this inter-pixel timing difference contains sufficient information to reconstruct the incident track angle to within a few degrees. Conversely, accurate reconstruction of the cluster hit time requires correcting for the track angle.

This study demonstrates a path toward transforming conventional 4D tracking systems, which measure three spatial coordinates and time, into 6D tracking systems that additionally reconstruct the two angular components of the particle trajectory. This terminology is illustrated in Figure~\ref{fig:6d-tracking-intro}.

\begin{figure}[htbp]
\centering
\includesvg[width=.49\textwidth]{diagrams/6d-tracking-intro.svg}
\includesvg[width=.49\textwidth]{diagrams/motivation.svg}
\caption{Left: terminology for 6D tracking. Right: potential application of 6D tracking for track seeding. \cite{wiki-1,wiki-2} \label{fig:6d-tracking-intro}}
\end{figure}

The ability to measure track angles will be key for future collider detectors. In high-luminosity hadron colliders such as FCC-hh, track angles can facilitate track seeding and triggering, reducing the substantial combinatorial background in track reconstruction at extreme luminosities~\cite{MGarcia-Sciveres_2010, Kar_2025}. In a muon collider, track angles provide an additional handle to suppress beam-induced backgrounds. One approach uses double layers, as implemented in the HL-LHC Tracker upgrade for CMS~\cite{CERN-LHCC-2017-009}. However, leveraging the fast-timing information provided by LGAD sensors enables the determination of the track angle using a single detector layer, provided that the track traverses at least three pixels.

Prior studies have established that pixel trackers can provide information beyond two-dimensional hit positions. Early measurements with monolithic pixel detectors demonstrated that the elongated shape of charge clusters from tilted tracks could provide information about the track angle~\cite{KENNEY199459}. Subsequent studies showed that if the shape of a cluster and the track angle are known, it is possible to achieve spatial resolution beyond the binary-readout pitch$/\sqrt{12}$ limit~\cite{Wang2018}, while ATLAS studies demonstrated that effects such as $\delta$-ray production can broaden clusters and bias purely geometric reconstruction methods~\cite{Aad2013}. Recently, machine-learning approaches have enabled probabilistic single-layer reconstruction of hit positions and track angles directly from pixel charge patterns under realistic ASIC constraints~\cite{das2026onchipprobabilisticinferencechargedparticle}. Additionally, a qualitative relationship between the peak amplitude of the pixel current and the track angle has been observed~\cite{osti_1841398}. While these studies relied on the spatial distribution of charge across the pixel array, the present work investigates a complementary approach that exploits the temporal information contained in the signals produced by tilted tracks. This work thus represents a quantitative investigation of the correlation between multi-pixel timing measurements and track angles, as well as a systematic study of crosstalk and Landau fluctuation effects on inter-pixel timing distributions in LGAD arrays. This timing information can be used to determine the sign of the track angle, which charge-based methods cannot provide. It can also be used either as a standalone method for computing the track angle via a single linear equation, or in combination with pixel charge information to further improve the angular resolution.

Integrating angle measurements into a standard track-fitting framework, such as a Kalman filter, would require addressing several practical considerations. The angle, position, and timing measurements extracted from a given layer are derived from a common set of pixel hit times, and are therefore correlated with one another and with the discrete pixel-identification step that precedes angle reconstruction; a rigorous treatment would need to account for the resulting off-diagonal measurement covariance rather than treating the angle as an independent scalar measurement. Additionally, angle reconstruction is unavailable for low track angles because an insufficient number of pixels is hit, and for some tracks, crosstalk between pixels strongly distorts the current waveforms and hence the reconstructed angle. The total cluster charge can be used to estimate the probability that a track is affected by such distortions, which could be propagated into the measurement covariance of a Kalman filter; in this paper we instead apply a hard selection cut to remove these tracks, and leave a complete treatment of this soft-weighting approach to future work.

This paper is organized as follows. Section~\ref{sec:lgads} provides an overview of LGAD detector physics and the TCAD modeling environment employed in this study. Section~\ref{sec:charge-theory} reviews existing charge-based methods for angle reconstruction and shows that they cannot resolve the sign of the track angle. Section~\ref{sec:timing-theory} derives an analytic relationship between inter-pixel timing differences and track incidence angles. This model fails to hold for a subset of tracks, which Section~\ref{sec:waveform-distortion} attributes to inter-pixel crosstalk driven by lateral track offset and by large Landau fluctuations in energy deposition. Section~\ref{sec:reconstruction} presents an algorithm for reconstructing the track angle from timing information, addressing the sign ambiguity left unresolved by the charge-based approach and correcting for the residual bias of the timing-based estimator. Section~\ref{sec:model-results} evaluates the performance of this algorithm, including a cluster-charge selection that rejects tracks affected by large Landau fluctuations, and compares it with a baseline that infers the angle from the cluster size alone. Section~\ref{sec:nn-models} compares these results with neural-network approaches that additionally exploit nonlinear correlations and pixel charge information. Appendices~\ref{sec:detailed-xtalk}--\ref{sec:nn-training} present a detailed analysis of crosstalk and linear equalization, the bias correction of the linear estimator, the effect of sensor thickness on angular reconstruction, and the neural-network training procedure.

\FloatBarrier
\section{The physics of LGAD detectors}
\label{sec:lgads}

LGADs \cite{FIRST-LGAD-PAPER} are silicon diode detectors incorporating an internal gain layer that provides signal amplification through charge multiplication. The structure of an LGAD is illustrated in Figure~\ref{fig:lgad-physics}~(a). The region between the p-type and n-type layers is fully depleted, establishing a high electric field that causes the generated electrons and holes to drift at close to their saturation velocities. Adjacent to the n-region, a high-dose implant further enhances the electric field magnitude, creating a gain layer. Electrons entering this gain region undergo impact ionization, initiating avalanche multiplication that results in current amplification. This charge multiplication is controlled by the applied bias voltage. Since high-gain operation requires bias voltages approaching the breakdown threshold, practical implementations typically maintain gain below $30\times$ to ensure stable operation~\cite{9081916}.

\begin{figure}[htbp]
\centering
\includesvg[width=.8\textwidth]{diagrams/lgad-physics.svg}
\caption{Operation of an LGAD detector following a hit by a high-energy particle: (a)~initial structure; (b)--(c)~creation of electron-hole pairs following interaction with the high-energy particle; (d)~drift of initial electron-hole pairs; (e)~multiplication of deposited electrons; (f)~drift and collection of multiplied charge carriers.\label{fig:lgad-physics}}
\end{figure}

When a charged high-energy particle traverses a semiconductor, the electric fields within the semiconductor lattice interact with the particle, causing energy loss. A portion of this deposited energy generates electron-hole pairs within the semiconductor, creating a trail of ionization along the particle trajectory, as shown in Figure~\ref{fig:lgad-physics}~(b) and (c). Following pair production, the internal drift field of the diode separates the charge carriers, driving electrons toward the n-side and holes toward the p-side, as shown in Figure~\ref{fig:lgad-physics}~(d). When electrons reach the gain region, they induce avalanche multiplication that produces additional electron-hole pairs, amplifying the current within the LGAD, as shown in Figure~\ref{fig:lgad-physics}~(e). Finally, both the primary and avalanche-generated charge carriers are collected at the diode terminals, as shown in Figure~\ref{fig:lgad-physics}~(f).

The LGAD geometry significantly affects detector performance. The sensor thickness is known to influence timing resolution~\cite{Berry:2022und}. However, as will be demonstrated, it also affects angular sensitivity, which requires that tilted tracks deposit charge across multiple pixels; this becomes more likely as the ratio of pixel pitch to sensor thickness decreases. The choice of sensor thickness should therefore account for both timing and angular performance, as discussed further in Appendix~\ref{sec:exploring-alternate-geometries}.

\FloatBarrier
\subsection{LGAD simulation environment and timing digitization}

\begin{figure}[htbp]
\centering
\includesvg[width=.5\textwidth]{diagrams/simulation-model.svg}
\\
\includesvg[width=.49\textwidth]{plots/energy-loss-sprectrum.svg}
\includegraphics[width=.49\textwidth]{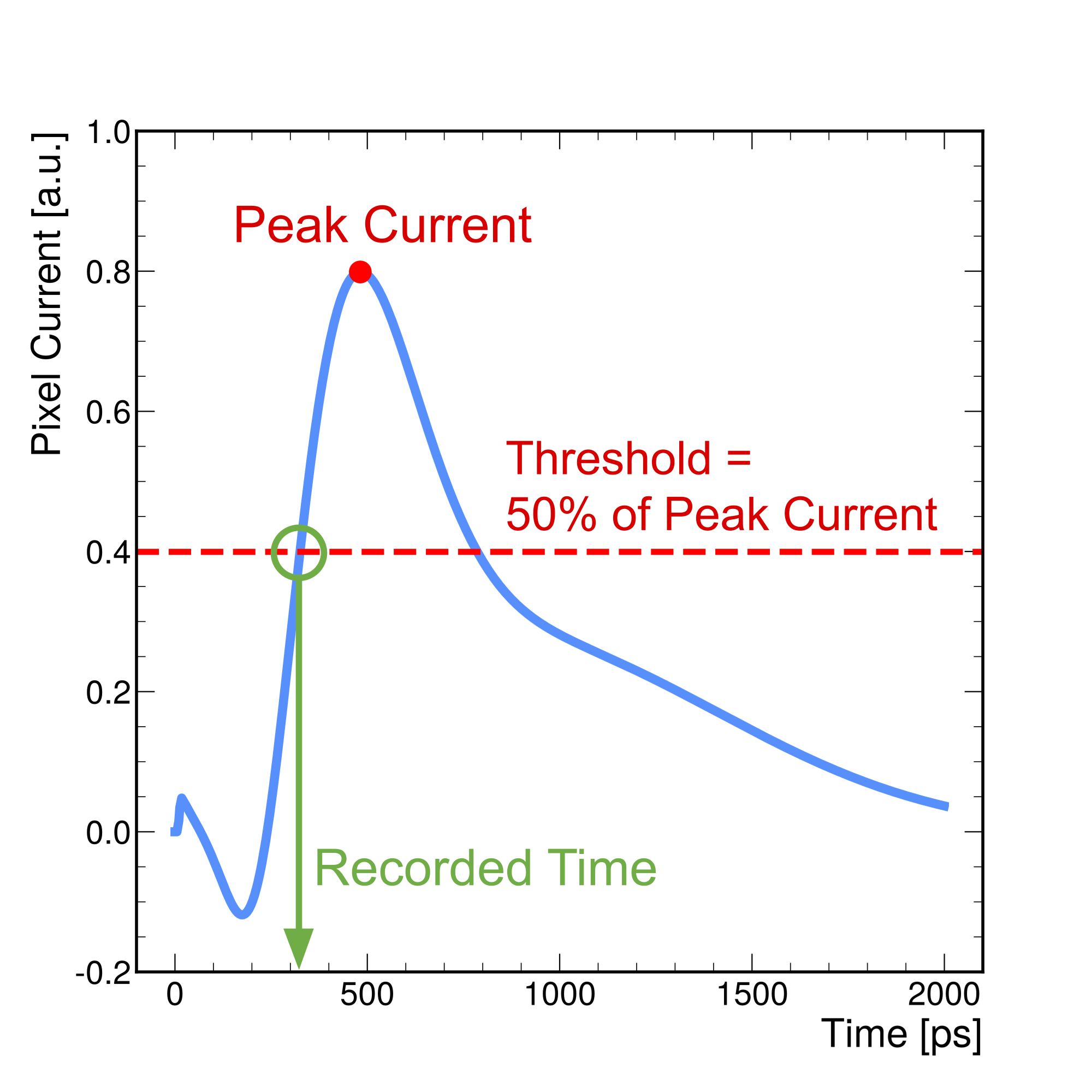}
\caption{Top: LGAD structure used throughout this study, shown with a $45^\circ$ track traversing the sensor. Bottom left: energy loss spectrum used to model Landau fluctuations. Bottom right: illustration of the CFD timing digitization method.\label{fig:simulation-model}}
\end{figure}

A comprehensive survey of LGADs can be found in Ref.~\cite{Berry:2022und}. The study in the present paper employs a trench-isolated LGAD design~\cite{9081916}, modeled using two-dimensional TCAD simulations with the Synopsys Sentaurus device simulator~\cite{Sentaurus}; the simulated device structure is shown in Figure~\ref{fig:simulation-model}. This two-dimensional treatment models a single row of pixels, corresponding to an LGAD strip detector or to tracks tilted purely along one axis of a pixel grid. Since the underlying physical mechanisms, namely electron drift to the gain region and inter-pixel crosstalk, are not specific to the simulated geometry, we expect the qualitative results to apply to full three-dimensional pixel arrays and other LGAD architectures, although confirming this requires further study. Effects specific to two-dimensional pixel grids, such as the generalization of the sign-reconstruction algorithm to arbitrary azimuthal track directions, are not addressed here.

The structure represents a simplified device with basic geometric features to improve computational efficiency. It consists of five pixels with a pitch $p$ of \qty{25}{\micro\meter} and a thickness $T$ of \qty{75}{\micro\meter}, with trench isolation providing inter-pixel segmentation. This structure represents an inner cross-section of a larger LGAD array; consequently, it does not include the lateral termination structures that would be present at the array periphery. The gain region is located on the pixelated side, and the device is biased to achieve a current gain of $10\times$. The resulting single-pixel timing resolution depends on the track angle, lateral offset and digitization scheme. In the present study, for a $45^\circ$ track, the central pixels achieve a timing resolution of approximately \qtyrange{20}{25}{\pico\second}, whereas the resolution degrades substantially for pixels with little deposited charge.

Since electron-hole pair generation is a stochastic process, we study the effects of Landau fluctuations using Monte Carlo simulations. For each particle trajectory, an ensemble of 110 tracks is generated with randomized energy deposition. In previous studies~\cite{SWARTZ200388, CENNA2015149}, this has been implemented by randomly placing individual electron-hole pairs and calculating the trajectory of each charge carrier as it traverses the LGAD. The Sentaurus device simulator~\cite{Sentaurus} does not support individual particle tracking; instead, it operates by calculating the density of electrons and holes throughout the device. In the present study, each track is therefore discretized into \qty{4}{\micro\meter} segments, with the electron-hole pair density in each segment set to a random sample of the energy loss spectrum shown in Figure~\ref{fig:simulation-model}~(bottom left). This spectrum is obtained by fitting a Landau distribution\footnote{For very short segments below \qty{1}{\micro\meter}, the energy loss spectrum cannot be approximated by a Landau distribution due to the appearance of resonance peaks~\cite{SuDongLandau}.} to curve $a$\footnote{This curve has a most probable total energy loss of $\Delta_p = 1.857~\textrm{keV}$ and a full width at half maximum of $w = 1.758~\textrm{keV}$ for a \qty{10}{\micro\meter} segment; these values are rescaled to the \qty{4}{\micro\meter} segment length used in this study as follows. The Landau distribution can be parameterized by a location parameter $\mu$ and a scale parameter $\sigma$ (Ref.~\cite{reference.wolfram_2026_landaudistribution}), for which the sum of two independent Landau-distributed variables $X_1+X_2$ is itself Landau-distributed with location parameter $\mu_1+\mu_2$ and scale parameter $\sigma_1+\sigma_2$. Since a \qty{10}{\micro\meter} segment can be treated as the sum of $10/4$ statistically independent \qty{4}{\micro\meter} segments (by the infinite divisibility of the Landau distribution), the location $\mu_4$ and scale $\sigma_4$ parameters for a single \qty{4}{\micro\meter} segment are obtained by first finding the parameters $\mu_{10}$ and $\sigma_{10}$ that reproduce $\Delta_p$ and $w$ for the \qty{10}{\micro\meter} segment, then scaling by $\mu_4 = \frac{4}{10}\mu_{10}$ and $\sigma_4 = \frac{4}{10}\sigma_{10}$.} in Figure 11 of Ref.~\cite{bichsel1988}, which corresponds to the energy loss spectrum of a 1 GeV electron in silicon. Secondary $\delta$-ray tracks are not modeled, since they occur rarely and are typically handled during track reconstruction. Similarly, following Appendix I of Ref.~\cite{bichsel1988}, bremsstrahlung need not be modeled to obtain realistic deposited-charge distributions for high-energy particles: although GeV-scale electrons produce substantial bremsstrahlung, it consists of high-energy X-rays with centimeter-scale absorption lengths, and therefore does not interact appreciably with the thin LGADs studied in this paper.

The process of analyzing a current waveform to determine a hit time is referred to as digitization. Multiple digitization schemes exist for fast-timing detectors; however, all must account for amplitude variations induced by Landau fluctuations. This paper adopts the Constant Fraction Discriminator (CFD) method, a common approach in timing applications. A CFD operates by creating two copies of the current pulse, one of which is delayed. The non-delayed copy is used to measure the pulse amplitude. The hit time is then defined as the instant when the delayed copy reaches a fixed fraction of the peak amplitude, which was determined using the non-delayed copy. For example, Figure~\ref{fig:simulation-model}~(bottom right) illustrates the operation of a CFD with a $50\%$ threshold.

CFD implementation remains an active area of research~\cite{XIE2023168655}. This study employs an ideal CFD with a $50\%$ threshold for digitization. Non-ideal effects such as bandwidth limitations or signal integration in the readout electronics were not modeled. However, thermal noise contributions from the LGAD sensor and front-end electronics were considered by adding Gaussian noise with a standard deviation of \qty{10}{\pico\second} to the CFD output.

\FloatBarrier
\section{Using charge information to measure angles for uniform tracks}
\label{sec:charge-theory}

Existing methods estimate the angle of a track from the lateral spread and distribution of the deposited charge cluster~\cite{KENNEY199459}. Throughout this paper, the track angle $\lambda$ is measured from the sensor normal and is positive when the track crosses the gain region in the leftmost pixel of the cluster. As illustrated in Figure~\ref{fig:charge-width}, a track traversing the detector at increasingly large $|\lambda|$ crosses a correspondingly larger number of pixels. By comparing the relative amount of charge shared between the pixels of a cluster, it is possible to estimate the lateral extent $W$ of the charge cluster, and hence $|\lambda|$, with a precision finer than the width of a single pixel~\cite{Wang2018}.

\begin{figure}[htbp]
\centering
\includesvg[width=\textwidth]{diagrams/charge-width-diagram.svg}
\caption{Charge distribution across the pixel array for a uniform track at two incidence angles. Top: tracks crossing at $34^\circ$ and $45^\circ$. Bottom: collected, uniform charge density of width $W$.\label{fig:charge-width}}
\end{figure}

However, as shown in Figure~\ref{fig:charge-sign}, the shape of a charge cluster cannot determine the sign of the track angle, since tracks of opposite sign produce identical charge distributions.

\begin{figure}[htbp]
\centering
\includesvg[width=0.5\textwidth]{diagrams/charge-width-sign-amibiguity.svg}
\caption{Charge distributions produced by tracks of equal magnitude but opposite sign. Top: tracks at $+45^\circ$ and $-45^\circ$ crossing the detector. Bottom: resulting identical collected charge densities.\label{fig:charge-sign}}
\end{figure}

\FloatBarrier
\section{Using timing information to measure angles for uniform tracks}
\label{sec:timing-theory}

In order to understand the relationship between timing and track angles, we consider a track traversing the LGAD at an oblique angle, as shown in Figure~\ref{fig:qualitative-diagram}. In the leftmost pixel, electrons are deposited next to the gain region and are immediately amplified. Conversely, in the rightmost pixel, electrons must first drift upward to reach the gain region. This drift across the full detector thickness can require hundreds of picoseconds. Consequently, the current waveform produced by the rightmost pixel is delayed relative to that of the leftmost pixel.

These delays are evident in the right-hand plot of Figure~\ref{fig:qualitative-diagram}: the current waveform of the left pixel peaks more than \qty{500}{\pico\second} before that of the right pixel. This has an immediate consequence for 4D tracking. When a track crosses multiple pixels, each pixel records a different hit time, so the particle arrival time cannot be obtained by simply averaging the hit times of the pixels in a cluster. Instead, the angle and lateral offset of the track must be accounted for in order to predict the expected timing variation between neighboring pixels. Whether the track angle can be determined in advance depends on the detector geometry and on the projection considered. In the $r$--$\phi$ projection of a barrel layer, the local track angle depends on the transverse momentum of the individual track. In the $r$--$z$ projection, the magnetic field does not change the track angle, but the longitudinal extent of the luminous region and multiple scattering in upstream material can still produce a substantial spread of angles at a given detector location. In general, the track angle and offset must therefore be treated as free parameters during track reconstruction, and angle compensation will be necessary to achieve the $\mathcal{O}(10~\mathrm{ps})$ timing resolution targeted by precision-timing detectors.

\begin{figure}[htbp]
\centering
\raisebox{1.1cm}{\includesvg[width=.49\textwidth]{diagrams/qualitative-diagram.svg}}
\hfill
\includesvg[width=.49\textwidth]{plots/landau_q_study_ratio_1.svg}
\caption{Left: diagram illustrating the link between timing and track angles, highlighting the difference in drift distances between neighboring pixels caused by tilted tracks. Right: current waveforms resulting from a $45^\circ$ tilted track.\label{fig:qualitative-diagram}} 
\end{figure}

A simple analytic model relating timing to track angles can be derived from two assumptions: (1) the readout electronics record the time corresponding to the leading edge of the current waveform, and (2) electrons travel at a constant saturation velocity within the LGAD.

As discussed in Section~\ref{sec:lgads}, electron-hole pairs begin to drift immediately upon creation due to the electric field within the LGAD, inducing a small current in the pixels. However, the current exhibits a significant peak only when electrons reach the gain region and undergo amplification. Since the readout electronics record the time of the leading edge of the pulse, the digitized time corresponds to the arrival of the first electrons at the gain region, rather than the initial ionization time.

Quantitatively, Figure~\ref{fig:linear-model-derivation}~(a) shows a typical case where a track traverses a pixel. The time recorded by that pixel is $t_i = t_{\text{hit}} + d_i/v_{\text{drift}}$, where $t_{\text{hit}}$ is the time when the particle intersects the detector, $d_i$ is the distance between the gain region and the point of the ionization track within pixel $i$ closest to it, and $v_{\text{drift}}$ is the electron saturation velocity within the LGAD.

This linear relationship can be exploited to deduce the track angle. Figure~\ref{fig:linear-model-derivation}~(b) shows a track traversing three pixels. Using timing information, we obtain $d_0=(t_0-t_{\text{hit}})v_{\text{drift}}$ and $d_1=(t_1-t_{\text{hit}})v_{\text{drift}}$. From geometry, $\Delta x=p$ (the pixel pitch), $\Delta y = d_1-d_0$, and $\cot \lambda = \Delta y/\Delta x$. Combining these equations yields Equation~\eqref{eq:cot-formula}.

\begin{equation}
\label{eq:cot-formula}
\begin{aligned}
    \cot \lambda=\frac{v_{\text{drift}}}{p}\left(t_1-t_0\right)
\end{aligned}
\end{equation}

\begin{figure}[htbp]
\centering
\includesvg[width=.8\textwidth]{diagrams/linear-model-derivation.svg}
\caption{Diagrams for deriving the relationship between timing and track angles. (a): illustration of the drift time required for an electron-hole pair to reach the gain region. (b): definition of geometric quantities used in deriving the linear model.\label{fig:linear-model-derivation}}
\end{figure}

Despite its simplicity, this linear model agrees well with full TCAD simulations, as shown in Figure~\ref{fig:cot-plot}.

\begin{figure}[htbp]
\centering
\includesvg[width=.5\textwidth]{plots/cot-plot.svg}
\caption{Comparison between the timing predictions of the linear model and device simulations. Fitting the linear model to the simulations yields $v_\text{drift} = 10^7~\mathrm{cm/s}$.\label{fig:cot-plot}}
\end{figure}

The linear model exhibits several notable properties. First, Equation~\eqref{eq:cot-formula} is independent of the lateral position of the track. For example, Figure~\ref{fig:edge-cases}~(a) shows two parallel tracks. Although each track produces different absolute times in both pixels due to their lateral offset, both tracks yield the same value of $\Delta y$. Consequently, the time difference between pixels is identical for both tracks and correctly determines the track angle. A second property is that the timing in the pixel where the track intersects the gain region is independent of track angle or lateral offset. Since electron-hole pairs are created within the gain region itself, they undergo immediate amplification without drift-induced, angle-dependent delay. For example, this occurs in the leftmost pixel of Figure~\ref{fig:linear-model-derivation}~(b), which yields $t_{-1}=t_{\text{hit}}$. Consequently, angle determination requires first identifying the track direction to locate this leading pixel, i.e., determining the sign of the angle. Furthermore, at least three pixels must be hit to unambiguously determine the track angle using timing information alone, since the leading pixel carries no angular information and a single remaining pixel cannot separate the effects of angle and lateral offset. Figure~\ref{fig:edge-cases}~(b) illustrates two tracks that traverse only two pixels. Despite having different angles, these tracks produce identical timing signatures.

\begin{figure}[htbp]
\centering
\includesvg[width=.8\textwidth]{diagrams/edge-cases.svg}
\caption{(a): illustration of the invariance of angular determination with respect to track lateral offset. (b): illustration of two tracks traversing only two pixels with identical timing information, demonstrating that the angle cannot be unambiguously reconstructed.\label{fig:edge-cases}}
\end{figure}

Finally, the linear model enables a first-order estimate of the achievable angular resolution for the studied LGAD. According to Figure~\ref{fig:cot-plot}, over the range from $30^\circ$ to $50^\circ$, the timing difference between two neighboring pixels varies by \qty{230}{\pico\second}. Taking a single-pixel timing resolution of approximately \qty{20}{\pico\second} for a $45^\circ$ track and electronic noise of \qty{10}{\pico\second}, and assuming that the timing errors of neighboring pixels are uncorrelated, the resolution of $t_1-t_0$ is approximately $\sqrt{2}\sqrt{(10~\mathrm{ps})^2+(20~\mathrm{ps})^2}\approx32~\mathrm{ps}$. The expected angular resolution from the timing information of two pixels is then $\sigma_\lambda\approx\frac{20^\circ}{230~\mathrm{ps}}\times 32~\mathrm{ps}\approx2.8^\circ$.

\FloatBarrier
\section{The distortion of current waveforms from crosstalk}
\label{sec:waveform-distortion}

When the track of a high-energy particle crosses multiple pixels, it induces current pulses in each pixel. For a uniform, centered track, this process is simple: each pixel produces a single current pulse, with inter-pixel timing delays set by the electron drift distances, as shown in Figure~\ref{fig:qualitative-diagram}~(right). Real tracks, however, are subject to two additional effects, lateral offset and Landau fluctuations, both of which can distort this simple picture.

The creation of an electron-hole pair within a pixel induces a current in that pixel through the drift and multiplication of the charge carriers. The motion of these carriers also induces smaller currents in neighboring pixels, a phenomenon known as crosstalk~\cite{10.1063/1.1710367}. Figure~\ref{fig:pix-response} shows simulated pixel current waveforms resulting from a localized charge deposit at the center of the pixel array: the current induced in the central pixel is significantly larger in amplitude than the crosstalk currents induced in its neighbors. Consequently, for tracks with a uniform charge distribution, each pixel's waveform is dominated by its own locally generated carriers, with crosstalk contributing only a small negative undershoot in the initial phase of the waveform; this has a negligible effect on peak timing or overall pulse shape, as seen in Figure~\ref{fig:qualitative-diagram}~(right).

\begin{figure}[htbp]
\centering
\includesvg[width=.5\textwidth]{plots/landau_q_study_ratio_-1.svg}
\caption{Pixel currents induced by a localized charge deposit within the center pixel. The current waveforms of the left and right pixels are identical.\label{fig:pix-response}}
\end{figure}

This behavior changes when the total deposited charge per pixel becomes non-uniform. If the charge deposited in one pixel significantly exceeds that of its neighbors, the crosstalk current it induces in those neighboring pixels can dominate over their own locally generated current. Such large, localized charge imbalances arise from two distinct mechanisms: the lateral offset of the track, which shifts the charge distribution asymmetrically across the pixel array, and Landau fluctuations, which produce stochastic, pixel-to-pixel variations in deposited charge.

The combined effect of these two mechanisms is shown in Figure~\ref{fig:noise-plots}. Figure~\ref{fig:noise-plots-a} shows typical waveforms, representative of roughly $90\%$ of tracks, in which each pixel produces a single current pulse with timing delays set by the track angle, and Landau fluctuations introduce only modest amplitude variation across pixels. Figure~\ref{fig:noise-plots-b} shows the remaining $10\%$ of waveforms, in which large Landau fluctuations or track offsets produce a strongly non-uniform charge deposition. The resulting crosstalk current induced in neighboring pixels is large enough to distort both the timing and shape of their waveforms, in some cases producing double-peaked pulses. These two mechanisms of non-uniform charge deposition are discussed in detail in the following subsections.

\begin{figure}[htbp]
\centering
\begin{subfigure}{\textwidth}
\includesvg[width=.24\textwidth]{plots/noise-plots-normal-1.svg}
\includesvg[width=.24\textwidth]{plots/noise-plots-normal-2.svg}
\includesvg[width=.24\textwidth]{plots/noise-plots-normal-3.svg}
\includesvg[width=.24\textwidth]{plots/noise-plots-normal-4.svg} 
\caption{Pixel currents produced by $45^\circ$ tracks with typical pixel charge uniformity.\label{fig:noise-plots-a}}\end{subfigure}

\begin{subfigure}{\textwidth}
\includesvg[width=.24\textwidth]{plots/noise-plots-anomalous-4.svg}
\includesvg[width=.24\textwidth]{plots/noise-plots-anomalous-1.svg}
\includesvg[width=.24\textwidth]{plots/noise-plots-anomalous-2.svg}
\includesvg[width=.24\textwidth]{plots/noise-plots-anomalous-3.svg}
\caption{Pixel currents produced by $45^\circ$ tracks with significant pixel charge non-uniformity.\label{fig:noise-plots-b}}
\end{subfigure}
\caption{Ensemble of pixel current waveforms produced by an LGAD traversed by $45^\circ$ tracks at different lateral offsets and with Landau fluctuations present.\label{fig:noise-plots}}
\end{figure}

A comprehensive analysis of crosstalk in LGADs, including an evaluation of potential filtering strategies, is presented in Appendix~\ref{sec:detailed-xtalk}.

\FloatBarrier
\subsection{Non-uniform pixel charge caused by track offset}

Even a uniform track can deposit different amounts of charge in different pixels as a consequence of its lateral offset alone. When a track crosses multiple pixels, the path length within the central pixels is fixed by the track angle, while the path lengths within the two edge pixels depend on the lateral position of the track. This is illustrated in Figure~\ref{fig:illustration-lateral-offset}: under a lateral offset, the left and center pixels retain the same amount of deposited charge, and the charge deposited in the right pixel is reduced.

This charge non-uniformity leads to waveform distortion via crosstalk, as shown in Figure~\ref{fig:plot-lateral-offset}. In the left panel, the track is mostly centered in the center pixel, and although the right pixel's waveform has a small amplitude, it remains substantially larger than the initial negative dip caused by crosstalk from the center pixel. In the right panel, the track is shifted toward the left pixel, reducing the charge deposited in the right pixel and correspondingly lowering the amplitude of its waveform. The initial negative dip, however, retains the same amplitude, since it originates from crosstalk with the center pixel, whose deposited charge is unaffected by the offset. As a result, the peak and dip amplitudes become comparable, producing a visibly distorted waveform shape.

\begin{figure}[htbp]
\centering
\includesvg[width=.8\textwidth]{diagrams/lateral-offset-amplitudes.svg}
\caption{Left: centered track crossing three pixels, depositing comparable amounts of charge in each. Right: laterally offset track depositing the same amount of charge in the two left pixels as before, but substantially less charge in the right pixel.\label{fig:illustration-lateral-offset}}
\end{figure}

\begin{figure}[htbp]
\centering
\includesvg[width=.49\textwidth]{plots/lateral-track-offset-6.0.svg}
\includesvg[width=.49\textwidth]{plots/lateral-track-offset-16.0.svg}
\caption{Normalized current waveforms produced by a uniform $45^\circ$ track. Left: track mostly centered in the center pixel. Right: track offset toward the left pixel, leaving substantially less charge in the right pixel.\label{fig:plot-lateral-offset}}
\end{figure}

\FloatBarrier
\subsection{Non-uniform pixel charge caused by Landau fluctuations}

When a charged high-energy particle traverses an LGAD, it loses energy through interactions with the silicon lattice, resulting in the creation of electron-hole pairs. Both the number of interactions per unit length and the number of electron-hole pairs created per interaction are stochastic processes~\cite{bichsel1988}. Consequently, the spatial distribution of deposited charge exhibits random fluctuations, known as Landau fluctuations. Figure~\ref{fig:regular-tracks-rho} illustrates this by showing an ensemble of possible ionization tracks generated by particles with identical trajectories.

\begin{figure}[htbp]
\centering
\includesvg[width=\textwidth]{diagrams/regular-tracks-rho.svg}
\caption{Ensemble of ionization tracks generated by particles with identical trajectories traversing at $45^\circ$ with Landau fluctuations present.\label{fig:regular-tracks-rho}}
\end{figure}

Landau fluctuations are characterized by a long-tailed distribution~\cite{bichsel1988}, which can be understood through a simple physical picture. As a high-energy particle traverses the detector, it experiences an average stopping force due to the electric fields produced by the silicon lattice. Occasionally, however, the particle undergoes a hard collision with atomic electrons that generates thousands of electron-hole pairs in a localized region. An example of such a large fluctuation is shown in Figure~\ref{fig:delta-track-rho}.

\begin{figure}[htbp]
\centering
\includesvg[width=\textwidth]{diagrams/delta-track-rho.svg}
\caption{Example of a large Landau fluctuation. Left: typical ionization track. Right: track with identical trajectory but with a large Landau fluctuation leading to a localized charge deposit near the bottom of the LGAD.\label{fig:delta-track-rho}}
\end{figure}

A well-known consequence of the long tail of the Landau distribution is the production of $\delta$-rays, which are commonly discussed in the tracking and clustering literature. These occur when an incident particle transfers enough energy to an atomic electron that the electron is ejected with significant kinetic energy. This energetic electron then traverses the detector as a secondary particle, creating an additional track of electron-hole pairs. While $\delta$-rays affect spatial reconstruction mainly through cluster broadening~\cite{Aad2013}, timing applications are significantly more sensitive to the non-uniformity in ionization density caused by long-tail events.

\begin{figure}[htbp]
\centering
\includegraphics[width=.8\textwidth]{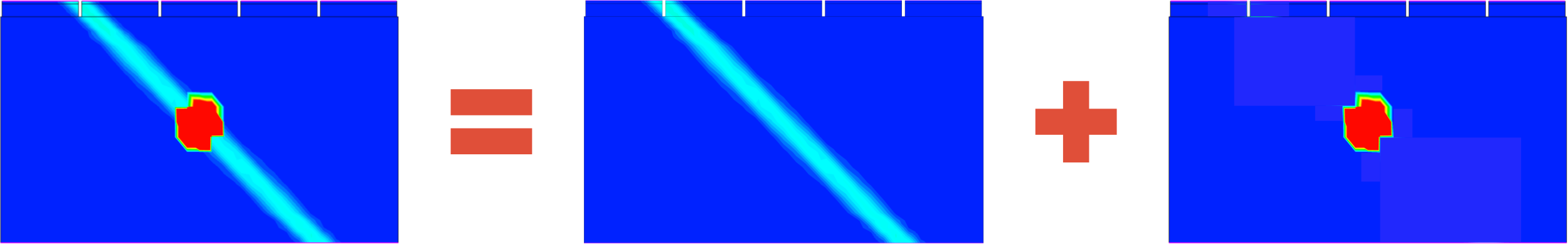}
\caption{Illustration of the model used to represent large Landau fluctuations. A track with a localized high-density charge cluster is decomposed into the sum of a uniform ionization track and an isolated charge cluster. \label{fig:linear-superposition}}
\end{figure}

\begin{figure}[htbp]
\centering
\includesvg[width=.49\textwidth]{plots/landau_q_study_ratio_5.svg}
\includesvg[width=.49\textwidth]{plots/landau_q_study_ratio_10.svg}
\\
\includesvg[width=.49\textwidth]{plots/landau_q_study_ratio_50.svg}
\includesvg[width=.49\textwidth]{plots/landau_q_study_ratio_200.svg}
\caption{Current waveforms produced by a $45^\circ$ track with a center-pixel charge cluster, as the cluster charge density is varied from small (uniform-track behavior) to large (isolated-cluster behavior). \label{fig:landau-q-sweep}}
\end{figure}

We model the effect of large Landau fluctuations by decomposing such a track into the linear superposition of a uniform track and a localized high-density charge cluster, as illustrated in Figure~\ref{fig:linear-superposition}. The resulting current waveforms are presented in Figure~\ref{fig:landau-q-sweep} for varying charge density ratios. Here, $n_\text{avg}$ denotes the electron-hole pair density along the uniform track component, $n_\delta$ represents the density within the localized cluster, and the ratio $n_\delta/n_\text{avg}$ is varied. For small values of $n_\delta/n_\text{avg}$, crosstalk contributions are negligible: the comparable charge deposition across all pixels yields waveforms similar to those of a uniform track, differing only in amplitude proportional to the local charge density. This regime corresponds to the behavior shown in Figure~\ref{fig:noise-plots-a}. As $n_\delta/n_\text{avg}$ increases, the substantial charge deposited within a single pixel induces significant crosstalk currents in neighboring pixels, giving rise to a second current peak. This behavior is consistent with the waveform characteristics observed in Figure~\ref{fig:noise-plots-b}.

\FloatBarrier
\section{Reconstructing track angles using the analytic models}
\label{sec:reconstruction}

Building on the analysis presented in the preceding sections, we now describe a simple algorithm for reconstructing the track angle. The determination of the track angle proceeds in two steps: first, the sign of the angle is determined, and second, its magnitude is determined. As discussed in the previous section, charge non-uniformity distorts the timing information extracted from a track, which can lead to significant errors in the reconstructed angle, most notably by causing the sign of the angle to be misidentified. We find that while the sign-reconstruction algorithm can be made robust to non-uniformity caused by lateral track offset, distortions caused by large Landau fluctuations are not fully recoverable. We therefore introduce a method to identify tracks that are likely affected by a large Landau fluctuation, so that such tracks can be flagged and filtered.

\FloatBarrier
\subsection{Determining the sign of the track angle}

The first step is to determine the sign of the angle, $\text{sgn}\,\lambda$. This is done by examining the covariance between pixel timing and pixel position: if, as pixel position $x_i$ increases (moving from left to right), the pixel timing $t_i$ increases as well, then the track angle is assigned a positive sign under our convention.

This yields a first approximation, $\widehat{\text{sgn}\,\lambda} = \text{sgn}\,\text{cov}\left(x_i, t_i\right)$. This model, however, fails to predict the correct sign for approximately $10\%$ of tracks (Figure~\ref{fig:sign-angle-det}), due to the waveform distortions discussed in Section~\ref{sec:waveform-distortion}. These distortions occur specifically when the charge deposited in a pixel is significantly smaller than that of its neighbors. We find that sign errors can be substantially reduced by weighting the covariance by the peak amplitude of each pixel's current waveform, which suppresses the contribution of low-amplitude pixels whose timing is dominated by crosstalk. The resulting algorithm proceeds as follows:

\begin{enumerate}
    \item Compute the averages weighted by the peak amplitude $A_i$ of the $i$th pixel:
    \begin{align}
        \bar{x} = \frac{\sum_i A_i x_i}{\sum_i A_i}, \qquad
        \bar{t} = \frac{\sum_i A_i t_i}{\sum_i A_i}
    \end{align}
    \item Compute the weighted covariance:
    \begin{align}
        C = \frac{\sum_i A_i \left(x_i - \bar{x}\right)\left(t_i - \bar{t}\right)}{\sum_i A_i}
    \end{align}
    \item Assign the sign of the track angle:
    \begin{align}
        \widehat{\text{sgn}\,\lambda} = \text{sgn}\, C
    \end{align}
\end{enumerate}

\subsection{Calculating the angle magnitude using timing information}

According to the linear model, the time difference between neighboring pixels is related to the track angle through Equation~\eqref{eq:cot-formula}, with the exception of the pixel in which the track intersects the gain region. Since the sign of the track angle is already known at this stage, we can construct a linear predictor for $\cot\lambda$:

\begin{equation}
    \widehat{\cot\lambda}_\text{biased}=c_0+\sum_i c_i\left(t_{i+1}-t_i\right)
\end{equation}

where pixels are ordered starting from the leading pixel along the track direction and the coefficients are fit separately for three-, four-, and five-pixel clusters. This estimator exhibits a small bias, linear in the true track angle, which can be removed through a simple numerical inversion, detailed in Appendix~\ref{sec:numerical-inversion}. The resulting bias-corrected estimator is:

\begin{equation}
\widehat{\lambda}=\widehat{\text{sgn}\,\lambda}\left[\alpha\cot^{-1}\left(c_0+\sum_i c_i\left(t_{i+1}-t_i\right)\right) + \beta\right]
\label{eq:estimator}
\end{equation}

Notably, this estimator is invariant under time translations, $t_i \rightarrow t_i + \tau$.

\FloatBarrier
\section{Track angle resolution using the analytic models}
\label{sec:model-results}

The performance of the estimator described in Section~\ref{sec:reconstruction} was evaluated using an ensemble of 56100 TCAD simulations with random Landau fluctuations. For each combination of track angle, ranging from $-50^\circ$ to $50^\circ$ in steps of $2^\circ$, and lateral offset, spanning ten values spaced \qty{2.5}{\micro\meter} apart, 110 independent simulations were generated. Of these, 16414 events hit three pixels, 10872 events hit four pixels, and 3800 events hit all five pixels. The linear coefficients of the estimator were fit using only tracks with positive angles, with tracks of negative angle serving as a validation set.

\subsection{Error rate and efficiency of sign determination}

The performance of the sign-reconstruction technique is shown in Figure~\ref{fig:sign-angle-det}~(left). The amplitude-weighted covariance technique is compared against an unweighted covariance method and a charge-weighted method, in which $A_i$ is replaced by $Q_i$, the total charge collected by pixel $i$. Performance is quantified by the cumulative sign error, defined as $\mathbb{P}\left(\text{error} \mid Q_\text{cluster} < x\right)$: the probability that a track exhibits a sign error given that its cluster charge, $Q_\text{cluster} = \sum_i Q_i$, is below $x$. The amplitude-weighted method outperforms both alternatives across the full range of cluster charge. Notably, considering clusters of all sizes, the amplitude-weighted method produces no sign errors for $Q_\text{cluster}$ below 2600~a.u., with the exception of a single outlier.\footnote{This outlier is a four-pixel track with a cluster charge of 2335~a.u., whose current waveforms were distorted by a rare Landau fluctuation that did not deposit a large total charge. It is removed from the dataset used in the remainder of this analysis. In a real system, such rare cases could instead be accommodated within a Kalman filter by retaining a small, nonzero probability of an incorrect sign even for low-charge tracks.} This is consistent with the idea that sign errors arise from large Landau fluctuations, which distort the pixel current waveforms by depositing unusually large amounts of charge along the track.

This relationship allows the probability of a sign error to be computed directly from the total cluster charge, which could, for instance, serve as an input to a Kalman filter. For simplicity, however, we instead adopt a hard cut, removing all tracks with cluster charge above 2600, 5400, and 6400~a.u. for three-, four-, and five-pixel tracks, respectively. Together with the removal of the single outlier, this eliminates sign errors entirely, enabling a clean study of the achieved angular resolution. Combined with the angle-consistency selection described in Section~\ref{sec:model-results}, these selections discard $8.9\%$ of eligible tracks. We note that these thresholds, and the resulting fraction of discarded tracks, are specific to the single-species, single-energy, background-free simulation studied here. In a collider environment, particles of different species and momenta deposit different average amounts of charge, so a single global threshold could reject ordinary tracks with higher $dE/dx$, such as slow, highly ionizing hadrons. Background radiation, secondaries, and detector effects such as gain non-uniformity and radiation damage may further reduce its effectiveness. The selection must therefore be studied under realistic conditions, potentially in combination with a Kalman filter whose momentum estimate informs the expected cluster charge.

The resulting efficiency of the algorithm is shown in Figure~\ref{fig:sign-angle-det}~(right). The efficiency is zero at low track angles, since angle reconstruction requires at least three pixels to be hit; it begins rising around $15^\circ$ and reaches $90\%$ at $30^\circ$, above which at least three pixels are hit for all tracks. Above $30^\circ$, the efficiency increases slightly further, as larger angles cause tracks to hit more pixels, providing additional data points and improving robustness against distortions.

\begin{figure}[!ht]
\centering
\includesvg[width=.49\textwidth]{plots/angle-sign-techniques.svg}
\includesvg[width=.49\textwidth]{plots/linear-model-efficiency.svg}
\caption{Top Left: histogram of the total collected charge for tracks with at least three pixels hit. Bottom Left: cumulative sign error for different sign-reconstruction algorithms, defined as $\mathbb{P}\left(\text{error}\mid Q_\text{cluster}<x\right)$. Right: reconstruction efficiency as a function of track angle, showing the fraction of tracks with at least three pixels hit that pass the cluster-charge selection.\label{fig:sign-angle-det}}
\end{figure}

\begin{figure}[htbp]
\centering
\includegraphics[width=.49\textwidth]{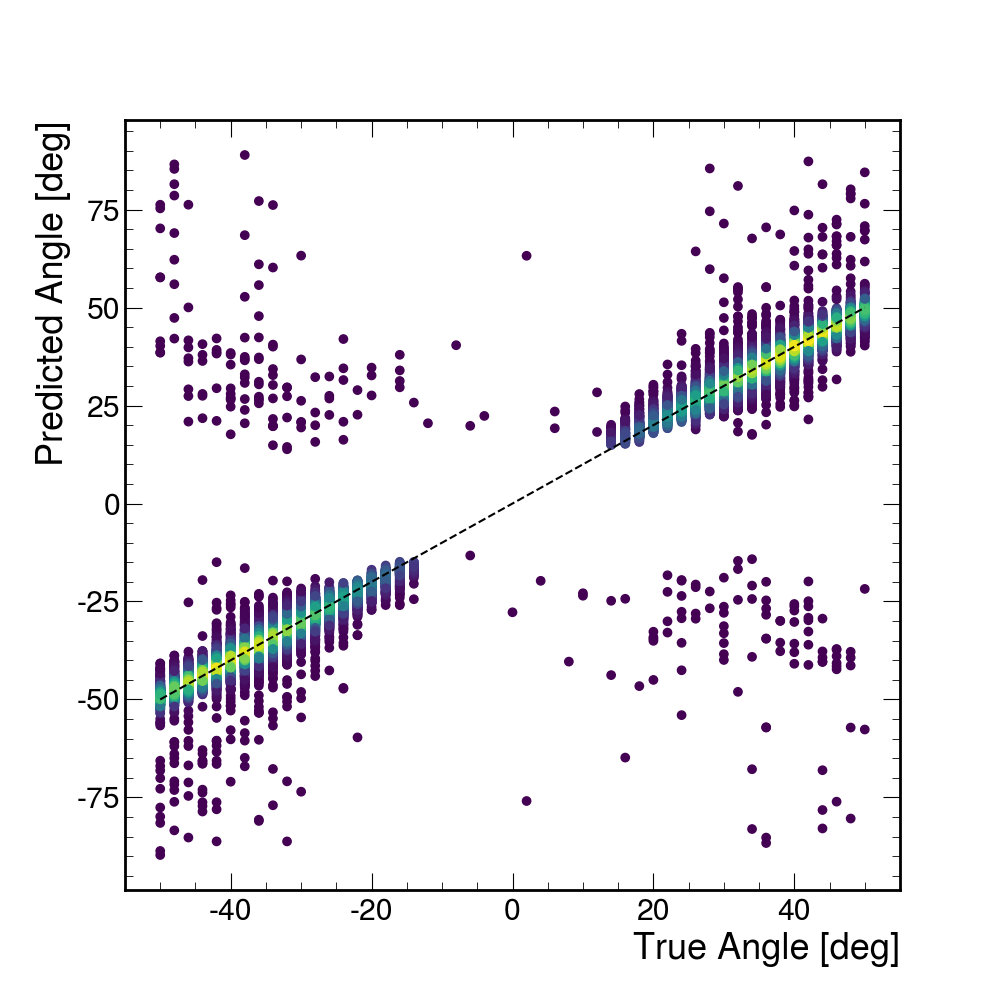}
\includegraphics[width=.49\textwidth]{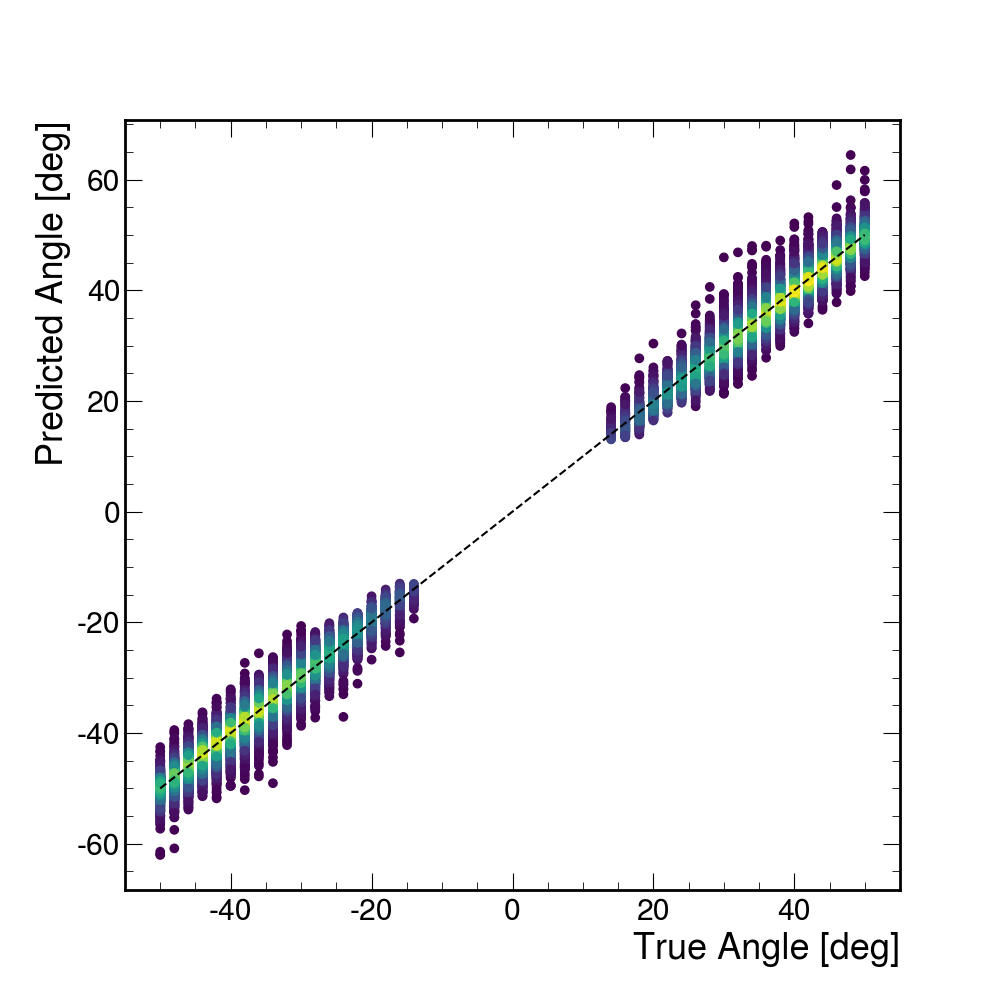}
\caption{Left: scatter plot of true vs.\ predicted angle for all tracks hitting at least three pixels, including the $0.54\%$ of tracks with sign errors. Right: scatter plot of true vs.\ predicted angle after charge filtering and the removal of geometrically inconsistent tracks.\label{fig:raw-scatter}}
\end{figure}

The effect of sign-determination errors is illustrated in Figure~\ref{fig:raw-scatter}. Although, with the amplitude-weighted method, such errors occur in only $0.54\%$ of tracks with at least three pixels hit, they introduce substantial distortions in the scatter plot and the resulting error distribution, justifying their removal when characterizing the resolution of the estimator.

\subsection{Angular magnitude resolution}

In addition to the cluster-charge selection, 96 four-pixel tracks were discarded because their reconstructed angles exceeded $57^\circ$, despite their signs being correctly reconstructed. Such angles are geometrically inconsistent with a track that hits only four pixels. In a real system, these cases could be handled naturally by a Kalman filter, whose prior on the track angle would assign negligible probability to such large angles. After all selections, 14255 tracks hitting three pixels, 10374 tracks hitting four pixels, and 3704 tracks hitting five pixels remain. The performance of the linear model for track angle reconstruction from timing information is evaluated by examining the distribution of the angular residual, defined as $\widehat{\lambda}_\text{res} = \widehat{\lambda}_\text{pred} - \lambda_\text{true}$, where $\lambda_\text{true}$ denotes the true track angle and $\widehat{\lambda}_\text{pred}$ denotes the angle reconstructed by the linear model. The angular bias and angular resolution are defined, respectively, as the mean and standard deviation of this distribution.

Figure~\ref{fig:linear-model-histograms} presents the angular residual distributions obtained from the linear model. For comparison, the residual distribution of a baseline approach, in which the track angle is inferred solely from the pixel cluster size, is also shown\footnote{The pixel-counting method assigns to each track the mean angle of all tracks traversing the same number of pixels. This method assumes the availability of an independent mechanism for determining the sign of the track angle.}. The linear model yields considerably narrower residual distributions than the pixel-counting approach, reflecting a substantial improvement in angular reconstruction performance. Furthermore, unlike the pixel-counting method, the residual distributions of the linear model are well described by a normal distribution.

\begin{figure}[!ht]
\centering
\includesvg[width=.32\textwidth]{plots/linear-model-error-histrograms-3.svg}
\includesvg[width=.32\textwidth]{plots/linear-model-error-histrograms-4.svg}
\includesvg[width=.32\textwidth]{plots/linear-model-error-histrograms-5.svg}
\caption{Angular residual distributions for three- (left), four- (center), and five-pixel (right) clusters using the linear model, compared with the pixel-counting baseline.\label{fig:linear-model-histograms}}
\end{figure}

The angular resolution and bias of the linear model are shown in Figure~\ref{fig:unbiased-model-resolution}, presented separately for three-, four-, and five-pixel clusters, as well as for all tracks combined. After numerical inversion, the angular bias remains below one degree across the full angular range studied. Clusters with a larger number of hit pixels achieve better angular resolution, as each additional pixel provides an additional timing measurement to constrain the model. Overall, the linear model achieves an angular resolution between two and three degrees across a wide range of incident angles. This is consistent with the first-order estimate of $\approx2.8^\circ$ derived in Section~\ref{sec:timing-theory}.

\begin{figure}[!ht]
\centering
\includesvg[width=.49\textwidth]{plots/linear-model-resolution-vs-angle-unbiased_algo_angle-3.svg}
\includesvg[width=.49\textwidth]{plots/linear-model-resolution-vs-angle-unbiased_algo_angle-4.svg}
\\  
\includesvg[width=.49\textwidth]{plots/linear-model-resolution-vs-angle-unbiased_algo_angle-5.svg}
\includesvg[width=.49\textwidth]{plots/linear-model-resolution-vs-angle-unbiased_algo_angle-3.4.5.svg}
\caption{Angular resolution and bias of the linear model as a function of track angle. Error bars indicate $68\%$ confidence intervals.\label{fig:unbiased-model-resolution}}
\end{figure}

In summary, a simple, physics-motivated estimator with few free parameters achieves an angular resolution below $3^\circ$ with a bias below $1^\circ$. We expect the qualitative results obtained in this section, particularly the near-linear relationship between timing and the cotangent of the track angle, and the distortion of current waveforms caused by crosstalk from non-uniform charge deposition, to generalize to other LGAD architectures and to pixel grids in which tracks may be tilted in an arbitrary azimuthal direction, although confirming this requires further study. A discussion of the effect of detector thickness on angular resolution may be found in Appendix~\ref{sec:exploring-alternate-geometries}.

\FloatBarrier
\section{Neural networks for angle reconstruction}
\label{sec:nn-models}

\begin{figure}[!ht]
\centering
\includegraphics[width=\textwidth]{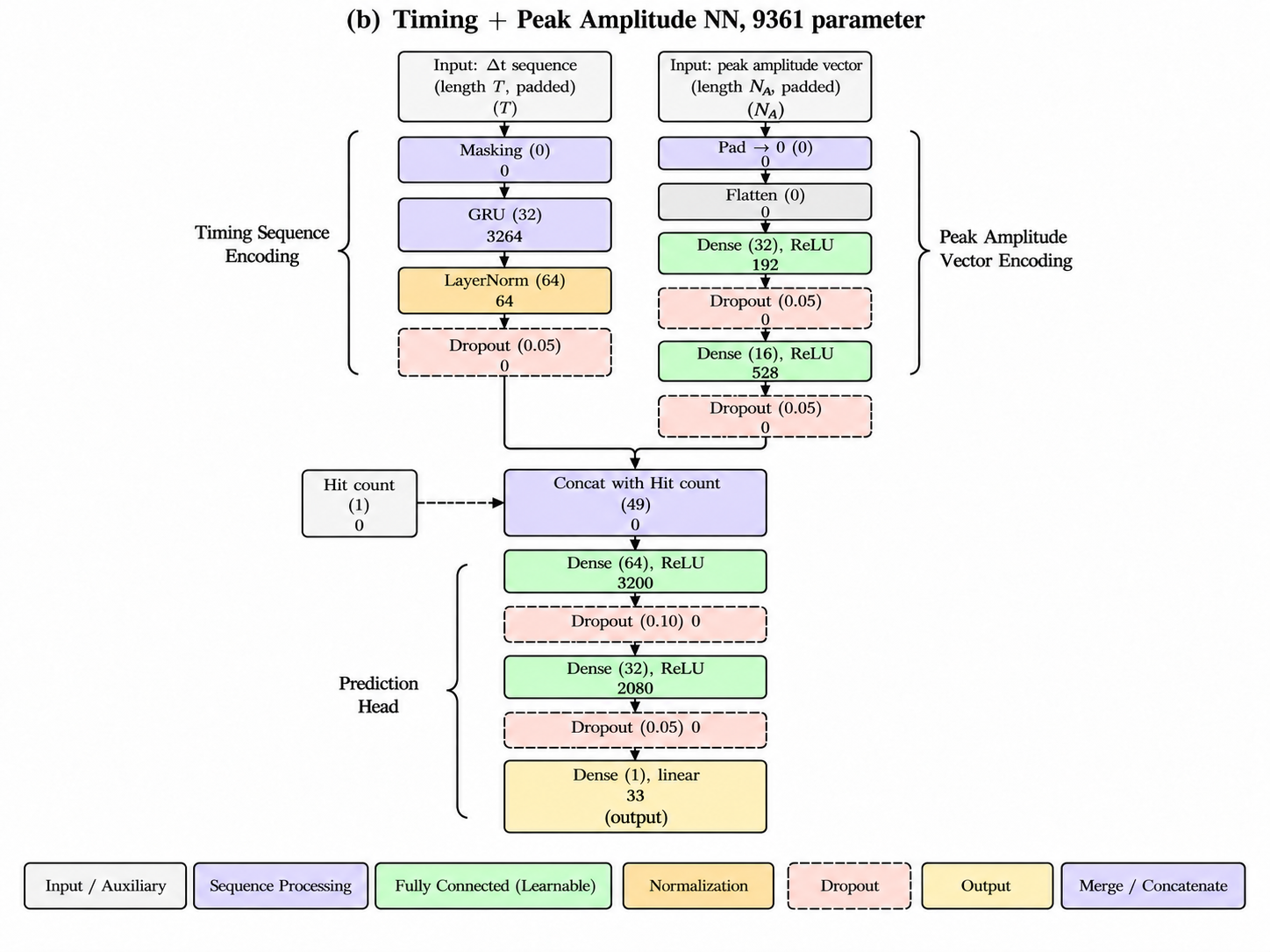}
\caption{Neural network architecture for the timing+amplitude model. The model takes as inputs the hit-time difference sequence ($\Delta t$) of length $T$, the peak amplitude vector of length $N_A$, and the hit multiplicity. The timing and amplitude inputs are processed by separate branches and subsequently concatenated with the hit multiplicity before the final regression layers. \label{fig:main_nn_architectures}}
\end{figure}

Unlike the linear model, a neural network can capture nonlinear relationships between timing and track angles.\footnote{An early example of using nonlinear functions of charge for track angle determination is found in Ref.~\cite{KENNEY199459}.} A neural network architecture is designed to reconstruct the incident track angle $\lambda$ from the hit times and peak amplitudes of the pixel current waveforms. The architecture is summarized in Figure~\ref{fig:main_nn_architectures}. Timing information is represented by consecutive hit-time differences $\Delta t_i = t_{i+1} - t_i$, consistent with the approach adopted in the linear model. 

The $\Delta t$ sequence is processed with a Gated Recurrent Unit (GRU) branch to model the ordered sequence of inter-pixel timing differences across the traversed pixels. The peak amplitude vector is processed with a dense neural network branch, which provides sufficient performance while maintaining architectural simplicity. The two learned representations, together with an explicit hit multiplicity, are concatenated and passed through a dense regression head, as shown in Figure~\ref{fig:main_nn_architectures}.

The parameter counts for this architecture are given in Figure~\ref{fig:main_nn_architectures}. Details of the neural network training procedure are given in Appendix~\ref{sec:nn-training}.

A second network, the timing+amplitude+charge NN, uses the same architecture, except that the peak-amplitude input is replaced by a vector of (amplitude, charge) pairs for each pixel.

The performance of each neural network model is summarized in Figure~\ref{fig:resolution_angle_comparison}. The timing+amplitude NN yields an angular resolution of approximately $2.0^\circ$, averaged over all angles, with bias below $1^\circ$ across the full angular range. The timing+amplitude+charge NN outperforms the timing+amplitude NN across the full angular range, yielding an angular resolution of approximately $1.7^\circ$, averaged over all angles, with bias below roughly $0.6^\circ$ across the full angular range. 

\begin{figure}[!ht]
    \centering
    \begin{minipage}{0.48\textwidth}
        \centering
        \includesvg[width=\textwidth]{sep30timeamplitude/standard_error_vs_angle_3-5_pixels_scipy_bootstrap_rnn_timeandcharge.svg}
    \end{minipage}
    \hfill
    \begin{minipage}{0.48\textwidth}
        \centering
        \includesvg[width=\textwidth]{sep30timechargeamplitude/standard_error_vs_angle_3-5_pixels_scipy_bootstrap_rnn_timechargeamplitude_combined.svg}
    \end{minipage}
    \caption{Angular resolution as a function of track angle. Left: timing+amplitude NN. Right: timing+amplitude+charge NN. Error bars indicate $68\%$ confidence intervals following numerical inversion.}
    \label{fig:resolution_angle_comparison}
\end{figure}

\begin{figure}[!ht]
    \centering
    \includegraphics[width=0.6\textwidth]{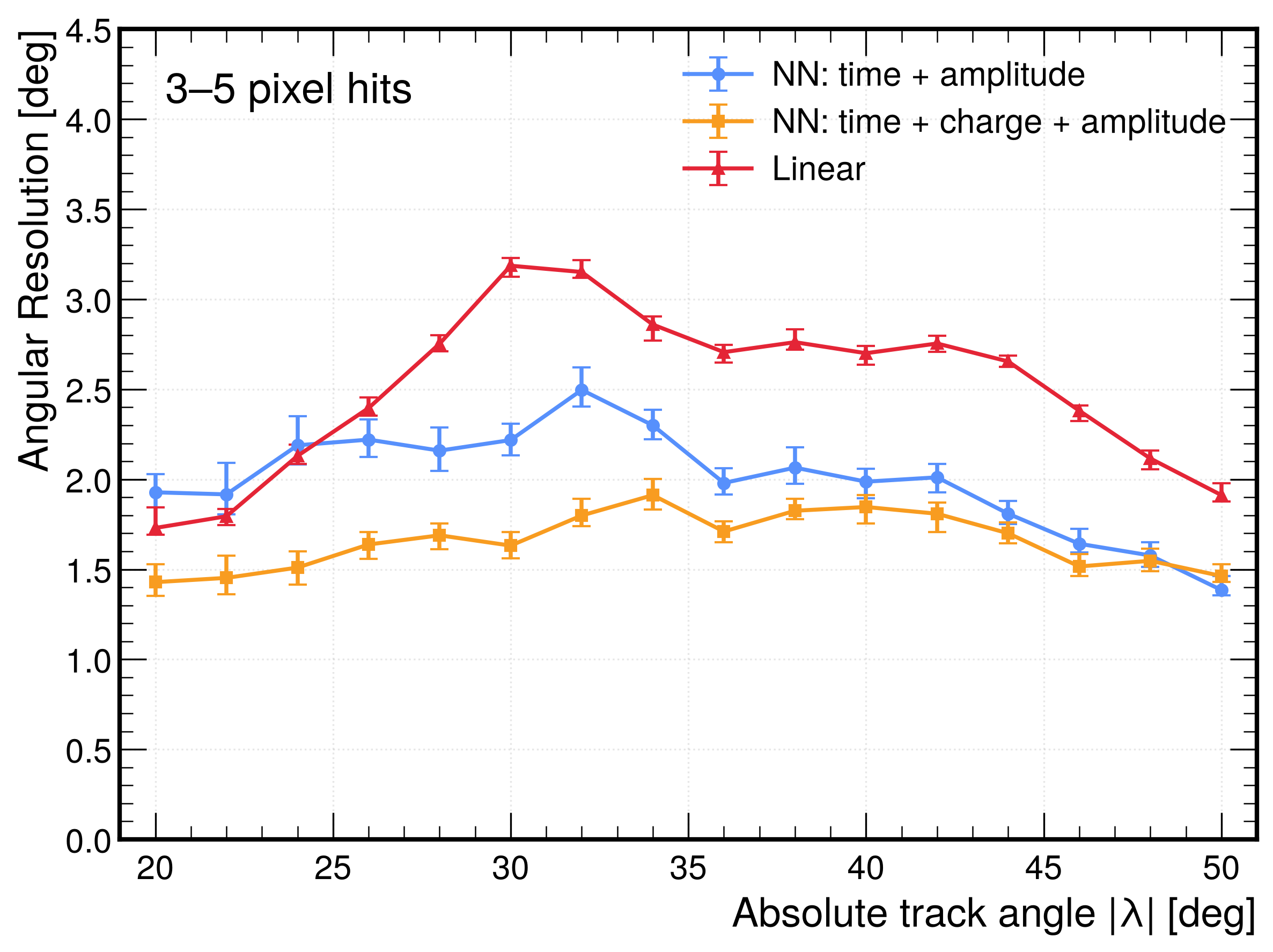}
    \caption{Angular resolution as a function of track angle for the timing+amplitude NN, the timing+amplitude+charge NN, and the linear model. Error bars indicate $68\%$ confidence intervals. Results are shown following numerical inversion.}
    \label{fig:resolution_angle_rnn_3to5linearvsnn}
\end{figure}

A comparison of the performance of the two neural networks and the linear model is shown in Figure~\ref{fig:resolution_angle_rnn_3to5linearvsnn}. As expected, both neural networks outperform the linear model, and the network with access to timing, charge, and amplitude information outperforms the network with access to timing and amplitude information alone. 

Above $25^\circ$, the network using timing and amplitude information outperforms the linear model by approximately $1^\circ$, while the network using timing, amplitude, and charge information outperforms the linear model by up to roughly $1.5^\circ$. At smaller angles, the network with access to charge information outperforms the other network by approximately $0.5^\circ$, but this advantage diminishes as the angle increases. 

\FloatBarrier
\section{Conclusion}

When a high-energy particle traverses multiple pixels of an LGAD-based 4D tracking detector, the track angle introduces timing variations of hundreds of picoseconds between pixels, even in detectors designed for tens-of-picoseconds timing resolution. This sensitivity has two important implications: first, an explicit correction for the track angle is necessary for accurate cluster hit time reconstruction; second, the track angle, including its sign (which cannot be obtained from the shape of the charge cluster), can be determined from timing information alone for tracks traversing at least three pixels.

The relationship between inter-pixel timing differences and the track angle follows a simple linear model, which enables angle reconstruction through a single linear combination of timing measurements. In two-dimensional TCAD simulations of a \qty{25}{\micro\meter} pitch, \qty{75}{\micro\meter} thick LGAD, this approach achieves an angular resolution of $2^\circ-3^\circ$ and a reconstruction efficiency of approximately $90\%$ above $\sim$30$^\circ$. A simple numerical inversion of the estimator reduces the angular bias to below $1^\circ$. Large Landau fluctuations are the dominant limitation: localized charge deposits induce crosstalk in neighboring pixels that distorts the recorded timing signals. Such tracks can be identified from their total cluster charge, and rejecting them removes nearly all sign errors. Together with the removal of a small number of geometrically inconsistent reconstructions, these selections discard $8.9\%$ of eligible tracks. Neural networks that additionally exploit nonlinear correlations and pixel charge improve the resolution by approximately $1^\circ-1.5^\circ$, to about $1.7^\circ$. The linear model nonetheless captures most of the available angular information at minimal computational cost, making it well-suited for implementation in on-chip or edge computing architectures.

Several steps remain before these techniques can be deployed, including extending the study to three-dimensional pixel arrays with arbitrary azimuthal track directions, incorporating realistic front-end electronics, integrating the angle measurement and its correlated uncertainties into track fitting, and validating the results with test-beam data. Timing-based angular measurements nonetheless provide a path toward 6D tracking systems, which measure position, time, and track angles in each tracker layer. Such capabilities could enable applications in track seeding, triggering, and background suppression at future high-luminosity colliders and other demanding experimental environments.

\appendix
\section{Detailed analysis of pixel crosstalk}
\label{sec:detailed-xtalk}

Crosstalk is a ubiquitous phenomenon in engineering systems, particularly in data transmission applications. This appendix provides a detailed analysis of crosstalk in LGADs and evaluates the feasibility of its mitigation through filtering techniques.

\begin{figure}[htbp]
\centering
\includegraphics[width=.2\textwidth]{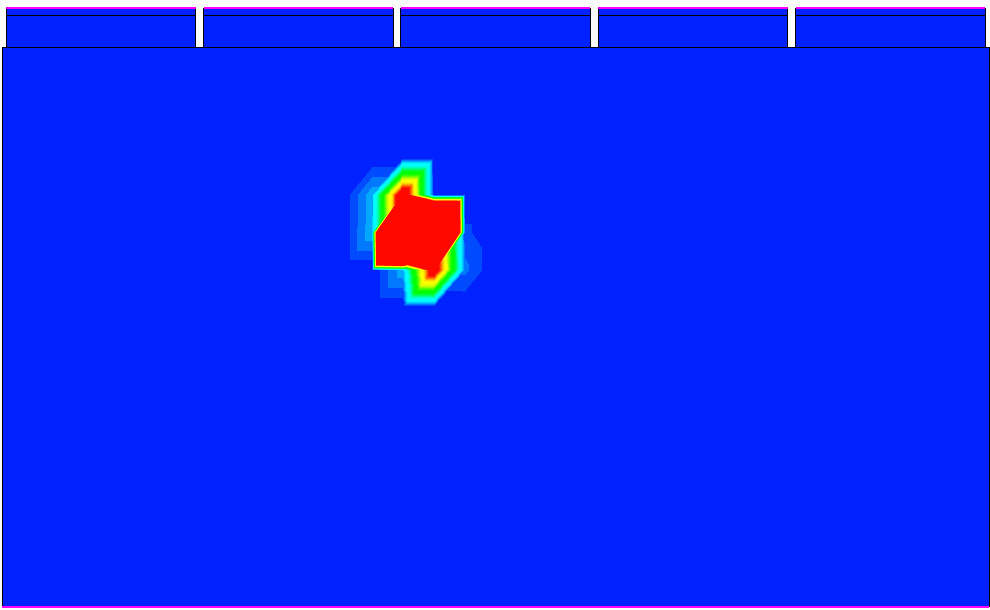}
\qquad
\includegraphics[width=.2\textwidth]{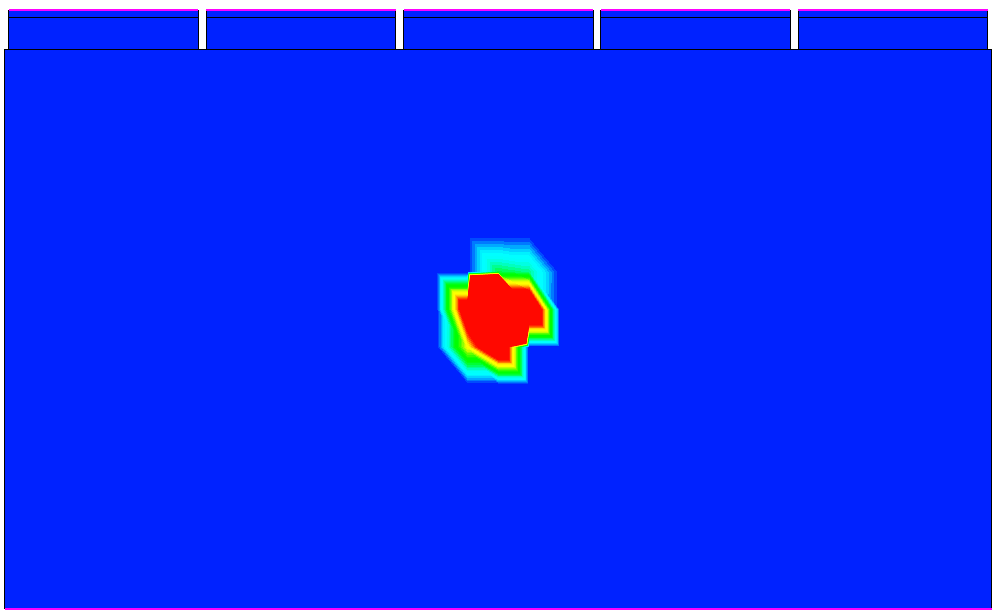}
\qquad
\includegraphics[width=.2\textwidth]{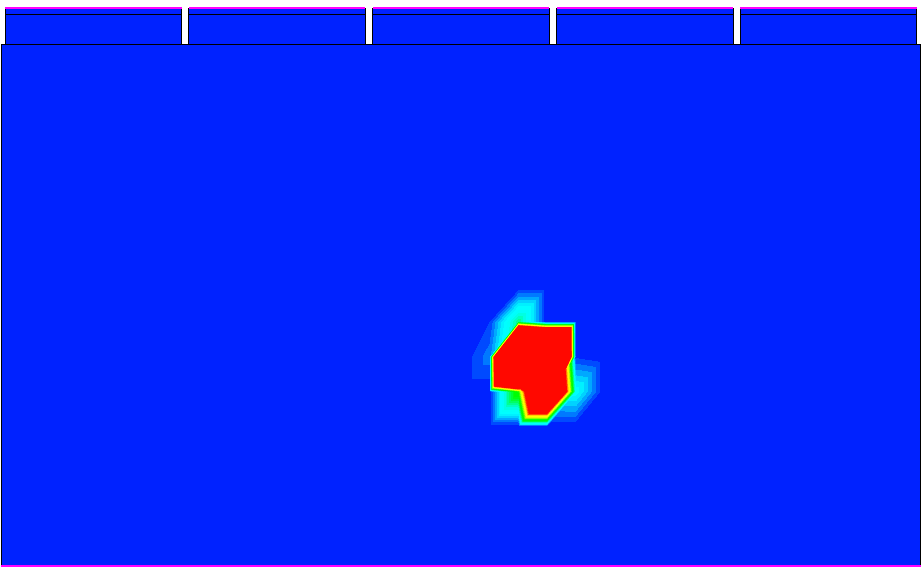}
\caption{Illustration of identical charge clusters placed at different lateral positions within a pixel to study induced crosstalk currents. Left: cluster positioned \qty{10}{\micro\meter} left of pixel center. Center: cluster at pixel center. Right: cluster positioned \qty{10}{\micro\meter} right of pixel center. \label{fig:sentaurus-lateral-offsets}}
\end{figure}

To characterize the crosstalk transfer function, simulations are performed with identical charge clusters placed at different lateral positions within a pixel, as illustrated in Figure~\ref{fig:sentaurus-lateral-offsets}. The resulting current waveforms are shown in Figure~\ref{fig:lateral-offset-i-plots}. These results demonstrate that the magnitude of induced crosstalk depends strongly on the lateral position of the charge cluster within the pixel: charge clusters closer to the pixel boundary induce larger crosstalk currents in neighboring pixels. In other words, the crosstalk transfer function varies as a function of the charge cluster's lateral position.

Linear equalization is a standard crosstalk reduction technique for linear systems with fixed transfer functions and is widely employed in communication systems~\cite{equalization_chapter}. Figure~\ref{fig:equalizer-plots} shows the residual crosstalk current when an ideal linear equalizer is applied to suppress crosstalk contributions. Although the equalizer reduces the crosstalk amplitude, significant residual current remains. This occurs because the crosstalk amplitude depends on the unknown lateral position of the charge cluster; consequently, the equalizer cannot achieve perfect cancellation.

In summary, unlike conventional communication systems where crosstalk transfer functions are fixed, LGAD crosstalk exhibits strong dependence on charge cluster position and therefore cannot be completely eliminated using simple linear equalization techniques.

\begin{figure}[htbp]
\centering
\includesvg[width=.49\textwidth]{plots/later_offset_of_anomaly_5.0.svg}
\includesvg[width=.49\textwidth]{plots/later_offset_of_anomaly_10.0.svg}
\\
\includesvg[width=.49\textwidth]{plots/later_offset_of_anomaly_-5.0.svg}
\includesvg[width=.49\textwidth]{plots/later_offset_of_anomaly_-10.0.svg}
\caption{Current waveforms produced by identical charge clusters placed at different lateral positions along a $45^\circ$ track. The cluster positions are offset laterally with respect to the center of the central pixel.\label{fig:lateral-offset-i-plots}}
\end{figure}

\begin{figure}[htbp]
\centering
\includesvg[width=.45\textwidth]{plots/equalizer-outputs-offset-5.svg}
\includesvg[width=.45\textwidth]{plots/equalizer-outputs-offset-10.svg}
\\
\includesvg[width=.45\textwidth]{plots/equalizer-outputs-offset--5.svg}
\includesvg[width=.45\textwidth]{plots/equalizer-outputs-offset--10.svg}
\\
\includesvg[width=.45\textwidth]{plots/equalizer-outputs-offset-0.svg}
\caption{Current waveforms after ideal linear equalization designed to minimize crosstalk. The crosstalk is induced by identical charge clusters placed at different lateral positions along a $45^\circ$ track, with positions offset laterally with respect to the center of the central pixel.\label{fig:equalizer-plots}}
\end{figure}

\FloatBarrier
\section{Numerical inversion for the linear model}
\label{sec:numerical-inversion}

The angular resolution and bias of the linear model prior to numerical inversion are shown in Figure~\ref{fig:biased-model-resolution}, presented separately for three-, four-, and five-pixel clusters, as well as for all tracks combined.

For each of the three-, four-, and five-pixel cases, the angular bias is observed to depend linearly on the true angle, with the same slope for positive and negative angles. This behavior can be expressed as:

\begin{equation*}
    \mathbb{E}\left[\widehat{\lambda}_\text{pred,biased}\right] = \lambda_\text{true} + \left(a\lambda_\text{true} + b \times \text{sgn}\left(\lambda_\text{true}\right)\right)
\end{equation*}

The parameters $a$ and $b$ are obtained via a linear fit of the bias as a function of angle, yielding the bias-corrected estimator:

\begin{equation*}
    \widehat{\lambda}_\text{pred} = \frac{\widehat{\lambda}_\text{pred,biased} - b \times \text{sgn}\left(\widehat{\lambda}_\text{pred,biased}\right)}{1+a}
\end{equation*}

Substituting into Eq.~\eqref{eq:estimator}, the bias correction corresponds to $\alpha=\frac{1}{1+a}$ and $\beta=-\frac{b}{1+a}$.

\begin{figure}[!ht]
\centering
\includesvg[width=.49\textwidth]{plots/linear-model-resolution-vs-angle-algo_angle-3.svg}
\includesvg[width=.49\textwidth]{plots/linear-model-resolution-vs-angle-algo_angle-4.svg}
\\
\includesvg[width=.49\textwidth]{plots/linear-model-resolution-vs-angle-algo_angle-5.svg}
\includesvg[width=.49\textwidth]{plots/linear-model-resolution-vs-angle-algo_angle-3.4.5.svg}
\caption{Angular resolution and bias of the linear model as a function of track angle before bias correction. Error bars indicate $68\%$ confidence intervals. \label{fig:biased-model-resolution}}
\end{figure}

\FloatBarrier
\section{The effect of sensor thickness on angular resolution}
\label{sec:exploring-alternate-geometries}

In this appendix, we discuss the impact of sensor thickness on track angle reconstruction. We argue that increasing the sensor thickness while maintaining a fixed pixel pitch is likely to improve angle reconstruction, although a full study is required to confirm this.

First, as shown in Figure~\ref{fig:ideal-efficiency}~(left), increasing the thickness at a fixed pitch increases the reconstruction efficiency. Since angle reconstruction requires at least three pixels to be hit, a thicker sensor satisfies this condition at smaller track angles.

Second, a thicker sensor may also improve the angular resolution. Figure~\ref{fig:ideal-efficiency}~(right) compares an identical track traversing a thin sensor (dark gray) and a thick sensor (light gray). In the thin sensor, useful timing information is available from only two pixels, $t_1$ and $t_2$.\footnote{As discussed in Section~\ref{sec:timing-theory}, the leftmost pixel always yields the track hit time, independent of the track angle.} Increasing the thickness may degrade the timing resolution of individual pixels, for example by lengthening the current pulse. We expect this degradation to be modest, however: as discussed in Section~\ref{sec:waveform-distortion}, the dominant source of timing fluctuations is the non-uniformity of charge deposition caused by Landau fluctuations, and the portion of the track traversing pixels~1 and~2 is identical in both sensors.

Furthermore, the thick sensor provides additional timing information from pixels~3 and~4, $t_3$ and $t_4$. Even if the single-pixel timing resolution degrades slightly, these additional measurements may lead to an overall improvement in angular resolution and potentially the cluster hit time resolution. Quantifying this balance requires dedicated simulations of thicker sensor geometries, which are left to future work.

\begin{figure}[htbp]
\centering
\includesvg[width=.49\textwidth]{plots/ideal-efficiency.svg}
\includesvg[width=.49\textwidth]{diagrams/detector-thickness-improvements.svg}
\caption{Illustration of the potential benefits of thicker sensors for angle reconstruction. Left: reconstruction efficiency, assuming no filtering, as a function of the pitch-to-thickness ratio $p/T$. Right: comparison of an identical track crossing a thin and a thick sensor.\label{fig:ideal-efficiency}}
\end{figure}

These considerations suggest that thicker LGADs may be worth investigating for applications in which angle measurements are desired, although the effect of increased thickness on timing resolution must also be considered~\cite{Berry:2022und}. Moreover, while a thicker LGAD adds material in the path of the incident particle, the net material budget could still be reduced if it enables a double-layer angle-measurement system to be replaced with a single layer, since the particle would no longer need to cross an additional sensor, readout ASIC, and associated support and cooling structures.

\FloatBarrier
\section{Neural network training}
\label{sec:nn-training}

The neural networks use the same 28333 tracks as the linear model, after charge filtering and the removal of geometrically inconsistent tracks. These were randomly divided into 22666 training samples ($80\%$) and 5667 validation samples ($20\%$). The training sample was used to optimize the neural-network parameters, while the validation sample was used to evaluate the model performance.

Timing, pixel amplitude and pixel charge inputs are zero-padded to fixed length for tracks with 3--5 hit pixels. 

All inputs are normalized using training-sample statistics:

\begin{align*}
\Delta t &\rightarrow \frac{\Delta t - \mu_{\Delta t}}{\sigma_{\Delta t}}, \\ A &\rightarrow \frac{A - \mu_A}{\sigma_A},  \\ Q &\rightarrow \frac{Q - \mu_Q}{\sigma_Q},
\end{align*}

where $\mu_x$ and $\sigma_x$ denote the corresponding mean and standard deviation of the training data, computed excluding padding values. The target angle is optionally standardized during training.

All models are trained using the Adam optimizer. Both models employ mean-squared-error loss with an angle-dependent event weighting scheme,
\begin{equation*}
w = 1 + 2\left(\frac{|\lambda|}{\lambda_{\max}}\right)^2,
\end{equation*}
with $\lambda_{\max}=50^\circ$ to enhance performance at large incident angles.

\FloatBarrier
\acknowledgments

The authors thank Dr. Stephen Richardson and Prof. Su Dong for helpful discussions and feedback. The authors thank the anonymous reviewer for their constructive comments, which helped identify a minor bug in the simulations.

\bibliographystyle{JHEP}
\bibliography{biblio}

@article{bichsel1988,
  title = {Straggling in thin silicon detectors},
  author = {Bichsel, Hans},
  journal = {Rev. Mod. Phys.},
  volume = {60},
  number = {3},
  pages = {663--699},
  year = {1988},
  doi = {10.1103/RevModPhys.60.663}
}

@techreport{osti_1841398,
  author       = {Lipton, Ronald J.},
  title        = {A Double Sided LGAD-Based Detector Providing Timing, Position, and Track Angle Information},
  institution  = {Fermi National Accelerator Laboratory (FNAL)}, address      = {Batavia, IL (United States)},
  doi          = {10.2172/1841398},
  url          = {https://www.osti.gov/biblio/1841398},
  place        = {United States},
  year         = {2022},
  month        = {01}}

@article{XIE2023168655,
title = {Design and performance of the Fermilab Constant Fraction Discriminator ASIC},
journal = {Nuclear Instruments and Methods in Physics Research Section A: Accelerators, Spectrometers, Detectors and Associated Equipment},
volume = {1056},
pages = {168655},
year = {2023},
issn = {0168-9002},
doi = {https://doi.org/10.1016/j.nima.2023.168655},
url = {https://www.sciencedirect.com/science/article/pii/S0168900223006459},
author = {Si Xie and Artur Apresyan and Ryan Heller and Christopher Madrid and Irene Dutta and Aram Hayrapetyan and Sergey Los and Cristián Peña and Tom Zimmerman}
}

@article{KENNEY199459,
title = {A prototype monolithic pixel detector},
journal = {Nuclear Instruments and Methods in Physics Research Section A: Accelerators, Spectrometers, Detectors and Associated Equipment},
volume = {342},
number = {1},
pages = {59-77},
year = {1994},
issn = {0168-9002},
doi = {https://doi.org/10.1016/0168-9002(94)91411-7},
url = {https://www.sciencedirect.com/science/article/pii/0168900294914117},
author = {Christopher J. Kenney and Sherwood I. Parker and Vincent Z. Peterson and Walter J. Snoeys and James D. Plummer and  {Chye Huat Aw}}
}

@inproceedings{Berry:2022und,
    author = "Berry, Doug and others",
    title = "{4-Dimensional Trackers}",
    booktitle = "{Snowmass 2021}",
    eprint = "2203.13900",
    archivePrefix = "arXiv",
    primaryClass = "physics.ins-det",
    reportNumber = "FERMILAB-CONF-22-284-PPD",
    month = "3",
    year = "2022"
}

@misc{Sentaurus,
  title = {{Sentaurus Device} An Advanced Multidimensional (1D/2D/3D) Device Simulator},
  howpublished = {\url{https://www.synopsys.com/manufacturing/tcad/device-simulation/sentaurus-device.html}},
  note = {Accessed: 2026-05-14}
}

@article{SWARTZ200388,
title = {CMS pixel simulations},
journal = {Nuclear Instruments and Methods in Physics Research Section A: Accelerators, Spectrometers, Detectors and Associated Equipment},
volume = {511},
number = {1},
pages = {88-91},
year = {2003},
note = {Proceedings of the 11th International Workshop on Vertex Detectors},
issn = {0168-9002},
doi = {https://doi.org/10.1016/S0168-9002(03)01757-1},
url = {https://www.sciencedirect.com/science/article/pii/S0168900203017571},
author = {M Swartz}
}

@article{CENNA2015149,
title = {Weightfield2: A fast simulator for silicon and diamond solid state detector},
journal = {Nuclear Instruments and Methods in Physics Research Section A: Accelerators, Spectrometers, Detectors and Associated Equipment},
volume = {796},
pages = {149-153},
year = {2015},
note = {Proceedings of the 10th International Conference on Radiation Effects on Semiconductor Materials Detectors and Devices},
issn = {0168-9002},
doi = {https://doi.org/10.1016/j.nima.2015.04.015},
url = {https://www.sciencedirect.com/science/article/pii/S0168900215004842},
author = {Francesca Cenna and N. Cartiglia and M. Friedl and B. Kolbinger and H.F.-W. Sadrozinski and A. Seiden and Andriy Zatserklyaniy and Anton Zatserklyaniy}
}

@ARTICLE{9081916,
  author={Paternoster, G. and Borghi, G. and Boscardin, M. and Cartiglia, N. and Ferrero, M. and Ficorella, F. and Siviero, F. and Gola, A. and Bellutti, P.},
  journal={IEEE Electron Device Letters}, 
  title={Trench-Isolated Low Gain Avalanche Diodes (TI-LGADs)}, 
  year={2020},
  volume={41},
  number={6},
  pages={884-887},
  doi={10.1109/LED.2020.2991351}}

@misc{SuDongLandau,
  author={Dong Su},
  title = {Landau Fluctuations},
  howpublished = {\url{https://indico.cern.ch/event/523323/}},
}

@misc{wiki-1,
  author={Jorge Stolfi},
  title = {File:Coord system CA 0.svg},
  howpublished = {\url{https://commons.wikimedia.org/wiki/File:Coord_system_CA_0.svg}},
  note = {Accessed: 2026-05-18}
}

@misc{wiki-2,
  author={Andeggs},
  title = {File:3D Spherical.svg},
  howpublished = {\url{https://commons.wikimedia.org/wiki/File:3D_Spherical.svg}},
  note = {Accessed: 2026-05-18}
}

@article{FIRST-LGAD-PAPER,
title = {Technology developments and first measurements of Low Gain Avalanche Detectors (LGAD) for high energy physics applications},
journal = {Nuclear Instruments and Methods in Physics Research Section A: Accelerators, Spectrometers, Detectors and Associated Equipment},
volume = {765},
pages = {12-16},
year = {2014},
note = {HSTD-9 2013 - Proceedings of the 9th International "Hiroshima" Symposium on Development and Application of Semiconductor Tracking Detectors},
issn = {0168-9002},
doi = {https://doi.org/10.1016/j.nima.2014.06.008},
url = {https://www.sciencedirect.com/science/article/pii/S0168900214007128},
author = {G. Pellegrini and P. Fernández-Martínez and M. Baselga and C. Fleta and D. Flores and V Greco and S. Hidalgo and I. Mandić and G. Kramberger and D. Quirion and M. Ullan}
}

@article{Wang2018,
  author = {Fuyue Wang and Benjamin Nachman and Maurice Garcia-Sciveres},
  title = {Ultimate position resolution of pixel clusters with binary readout for particle tracking},
  journal = {Nuclear Instruments and Methods in Physics Research Section A},
  volume = {899},
  pages = {10--15},
  year = {2018},
  doi = {10.1016/j.nima.2018.04.053}
}

@article{Aad2013,
  author = {G. Aad and others},
  collaboration = {ATLAS Collaboration},
  title = {A measurement of secondary hard interactions in minimum bias events with the ATLAS detector},
  journal = {New Journal of Physics},
  volume = {15},
  pages = {033038},
  year = {2013},
  doi = {10.1088/1367-2630/15/3/033038}
}

@misc{das2026onchipprobabilisticinferencechargedparticle,
      title={On-chip probabilistic inference for charged-particle tracking at the sensor edge}, 
      author={Arghya Ranjan Das and David Jiang and Rachel Kovach-Fuentes and Shiqi Kuang and Ana Sofía Calle Muñoz and Danush Shekar and Jennet Dickinson and Giuseppe Di Guglielmo and Lindsey Gray and Mia Liu and Corrinne Mills and Mark S. Neubauer and Daniel Abadjiev and Anthony Badea and Doug Berry and Karri DiPetrillo and Farah Fahim and Abhijith Gandrakota and Harshul Gupta and James Hirschauer and Eliza Howard and Ron Lipton and Petar Maksimovic and Nick Manganelli and Benjamin Parpillon and Jannicke Pearkes and Ricardo Silvestre and Morris Swartz and Chinar Syal and Nhan Tran and Amit Trivedi and Keith Ulmer and Mohammad Abrar Wadud and Benjamin Weiss and Eric You},
      year={2026},
      eprint={2602.15946},
      archivePrefix={arXiv},
      primaryClass={physics.ins-det},
      url={https://arxiv.org/abs/2602.15946}, 
}

@techreport{CERN-LHCC-2020-007,
      title         = "{Technical Design Report:  A High-Granularity Timing
                       Detector for the ATLAS Phase-II Upgrade}",
      institution   = "CERN",
      collaboration = "ATLAS Collaboration",
      address       = "Geneva",
      reportNumber  = "CERN-LHCC-2020-007, ATLAS-TDR-031",
      month         = "Jun",
      year          = "2020",
      url           = "https://cds.cern.ch/record/2719855",
}

@article{Butler:2019rpu,
    author = "Butler, Joel N. and Tabarelli de Fatis, Tommaso",
    collaboration = "CMS",
    title = "{A MIP Timing Detector for the CMS Phase-2 Upgrade}",
    reportNumber = "CERN-LHCC-2019-003, CMS-TDR-020",
    year = "2019"
}

@misc{bartosik2022simulateddetectorperformancemuon,
      title={Simulated Detector Performance at the Muon Collider}, 
      author={Nazar Bartosik and Karol Krizka and Simone Pagan Griso and Chiara Aimè and Aram Apyan and Mohammed Attia Mahmoud and Alessandro Bertolin and Alessandro Braghieri and Laura Buonincontri and Simone Calzaferri and Massimo Casarsa and Luca Castelli and Maria Gabriella Catanesi and Francesco Giovanni Celiberto and Alessandro Cerri and Grigorios Chachamis and Anna Colaleo and Camilla Curatolo and Giacomo Da Molin and Sridhara Dasu and Dmitri Desinov and Haluk Denizli and Biagio Di Micco and Tommaso Dorigo and Filippo Errico and Anna Ferrari and Davide Fiorina and Luca Giambastiani and Alessio Gianelle and Carlo Giraldin and Matthew Herndon and Tova Ray Holmes and Sergo Jindariani and Georgios Krintiras and Lawrence Lee and Qiang Li and Ronald Lipton and S. Lomte and Kenneth Long and Donatella Lucchesi and Paola Mastrapasqua and Federico Meloni and Alessandro Montella and Federico Nardi and Nadia Pastrone and Antonello Pellecchia and Karolos Potamianos and Emilio Radicioni and Raffaella Radogna and Cristina Riccardi and Luciano Ristori and Paola Salvini and I. Sarra and Daniel Schulte and Abdulkadir Senol and Lorenzo Sestini and Federica Maria Simone and Rosa Simoniello and Anna Stamerra and Xiaohu Sun and Maximilian J Swiatlowski and Jian Tang and Emily Anne Thompson and Ilaria Vai and Marco Valente and Nicolo' Valle and Rosamaria Venditti and Piet Verwilligen and Hannsjorg Weber and Angela Zaza and Davide Zuliani},
      year={2022},
      eprint={2203.07964},
      archivePrefix={arXiv},
      primaryClass={hep-ex},
      url={https://arxiv.org/abs/2203.07964}, 
}

@article{FCC_hh_2019,
  author       = {Abada, A. and Abbrescia, M. and AbdusSalam, S. S. and others},
  title        = {FCC-hh: The Hadron Collider},
  journal      = {European Physical Journal Special Topics},
  volume       = {228},
  number       = {4},
  pages        = {755--1107},
  year         = {2019},
  doi          = {10.1140/epjst/e2019-900087-0},
  url          = {https://doi.org/10.1140/epjst/e2019-900087-0}
}

@article{MGarcia-Sciveres_2010,
doi = {10.1088/1748-0221/5/10/C10001},
url = {https://doi.org/10.1088/1748-0221/5/10/C10001},
year = {2010},
month = {oct},
publisher = {},
volume = {5},
number = {10},
pages = {C10001},
author = {M Garcia-Sciveres and M Gilchriese and C Haber and B Heinemann and T Mueller},
title = {System concepts for doublet tracking layers},
journal = {Journal of Instrumentation}
}

@techreport{CERN-LHCC-2017-009,
      collaboration = "CMS",
      title         = "{The Phase-2 Upgrade of the CMS Tracker}",
      institution   = "CERN",
      reportNumber  = "CERN-LHCC-2017-009, CMS-TDR-014",
      address       = "Geneva",
      year          = "2017",
      url           = "https://cds.cern.ch/record/2272264",
      doi           = "10.17181/CERN.QZ28.FLHW",
}

@article{Kar_2025,
   title={A Triplet Track Trigger for the FCC-hh to improve the measurement of Di-Higgs production and the Higgs self-coupling},
   volume={1072},
   ISSN={0168-9002},
   url={http://dx.doi.org/10.1016/j.nima.2024.170085},
   DOI={10.1016/j.nima.2024.170085},
   journal={Nuclear Instruments and Methods in Physics Research Section A: Accelerators, Spectrometers, Detectors and Associated Equipment},
   publisher={Elsevier BV},
   author={Kar, T. and Schöning, A.},
   year={2025},
   month=Mar, pages={170085} }

@inbook{equalization_chapter,
author={Hall, S.H. and Heck, H.L.},
publisher = {John Wiley \& Sons, Ltd},
isbn = {9780470423899},
title = {Equalization},
booktitle = {Advanced Signal Integrity for High‐Speed Digital Designs},
chapter = {12},
pages = {499-548},
doi = {https://doi.org/10.1002/9780470423899.ch12},
url = {https://onlinelibrary.wiley.com/doi/abs/10.1002/9780470423899.ch12},
eprint = {https://onlinelibrary.wiley.com/doi/pdf/10.1002/9780470423899.ch12},
year = {2009}
}

@misc{reference.wolfram_2026_landaudistribution, author="Wolfram Research", title="{LandauDistribution}", year="2016", howpublished="\url{https://reference.wolfram.com/language/ref/LandauDistribution.html}"
}

@article{10.1063/1.1710367,
    author = {Shockley, W.},
    title = {Currents to Conductors Induced by a Moving Point Charge},
    journal = {Journal of Applied Physics},
    volume = {9},
    number = {10},
    pages = {635-636},
    year = {1938},
    month = {10},
    issn = {0021-8979},
    doi = {10.1063/1.1710367},
    url = {https://doi.org/10.1063/1.1710367}
}

\end{document}